\documentclass[
reprint,
superscriptaddress,
amsmath,amssymb,
aps,
]{revtex4-1}

\usepackage{graphicx,epsf,xcolor}

\begin{document}

\title{The hysteresis-free behavior of perovskite solar cells from the perspective of the measurement conditions}

\author{George Alexandru Nemnes}
\email{nemnes@solid.fizica.unibuc.ro}
\affiliation{University of Bucharest, Faculty of Physics, Materials and Devices for Electronics and Optoelectronics Research Center, 077125 Magurele-Ilfov, Romania}
\affiliation{Horia Hulubei National Institute for Physics and Nuclear Engineering, 077126 Magurele-Ilfov, Romania}
\author{Cristina Besleaga}
\email{cristina.besleaga@infim.ro}
\affiliation{National Institute of Materials Physics, Magurele 077125, Ilfov, Romania}
\author{Andrei Gabriel Tomulescu}
\affiliation{National Institute of Materials Physics, Magurele 077125, Ilfov, Romania}
\affiliation{University of Bucharest, Faculty of Physics, 077125 Magurele-Ilfov, Romania}
\author{Lucia Nicoleta Leonat}
\affiliation{National Institute of Materials Physics, Magurele 077125, Ilfov, Romania}
\author{Viorica Stancu}
\affiliation{National Institute of Materials Physics, Magurele 077125, Ilfov, Romania}
\author{Mihalea Florea}
\affiliation{National Institute of Materials Physics, Magurele 077125, Ilfov, Romania}
\author{Andrei Manolescu}
\affiliation{School of Science and Engineering, Reykjavik University, Menntavegur 1, IS-101 Reykjavik, Iceland}
\author{Ioana Pintilie}
\affiliation{National Institute of Materials Physics, Magurele 077125, Ilfov, Romania}

\begin{abstract}
We investigate in how far the hysteresis-free behavior of perovskite
solar cells can be reproduced using particular pre-conditioning and
measurement conditions. As there are currently more and more reports of
perovskite solar cells without J-V hysteresis it is crucial to distinguish
between genuine performance improvements and measurement artifacts. We
focus on two of the parameters that influence the dynamic J-V scans,
namely the bias scan rate and and the bias poling voltage, and point out
measurement conditions for achieving a hysteresis-free behavior. In this
context we discuss the suitability of defining a hysteresis index (HI)
for the characterization of dynamic J-V scans. Using HI, aging effects are
also investigated, establishing a potential connection between the sample
degradation and the variation of the maximal hysteresis on one hand, and
relaxation time scale of the slow process, on the other hand. Pre-poling
induced recombinations effects are identified. In addition, our analysis
based on sample pre-biasing reveals potential indication regarding two
types of slow processes, with two different relaxation time scales,
which provides further insight regarding ionic migration.
\end{abstract}

\maketitle

\section{Introduction}

Perovskite solar cells (PSCs) witnessed an impressive increase in power conversion efficiency (PCE) in the past few years, to date reaching 23.3\%, a certified but still unstabilized value \cite{nrel}. In the view of this spectacular development, the field of PSCs is most rapidly advancing amongst solar cell technologies. However, to attain a marketable status the stability issues signaled since the early days still need to be addressed. The degradation of the PSCs is related to both intrinsic factors, like the quality of the perovskite layer, diffusion of metal atoms or ions at the back contact electrodes \cite{doi:10.1021/acs.jpclett.6b02375,XIONG201630}, and extrinsic factors, like moisture, air, illumination and bias conditions \cite{C6TA09202K,DAO2017229}. In this context, ion migration was experimentally detected, potentially causing irreversible chemical reactions, and also identified as a primary cause for the J-V hysteresis \cite{doi:10.1021/acs.jpclett.7b00975}. 

Since its discovery, a number of potential reasons have been proposed to explain the J-V hysteresis: giant dielectric constant of the perovskite layer \cite{doi:10.1021/jz5011169}, ferroelectricity \cite{doi:10.1021/jz502111u,doi:10.1021/jz502429u,doi:10.1063/1.4890246}, trapping and de-trapping \cite{doi:10.1021/acs.jpclett.5b01645}, accumulations of ions and photogenerated carriers \cite{doi:10.1021/acs.jpclett.7b00045} or ion induced modifications of charge collection and recombination \cite{C4EE03664F}. Although presently a rather broad consensus has been reached regarding the role of ion migration as a trigger process, the precise mechanisms behind the J-V hysteresis and their quantitative contributions are still under debate \cite{doi:10.1002/aenm.201702772}. For instance, explanations include a large accumulation capacitance under illumination, as opposed to ion modified charge collection, which may be further influenced by trapping.

The J-V hysteresis is a dynamic process, which depends both on the pre-conditioning and measurement conditions, but also on material properties such as the crystallite size in the perovskite layer and the interfaces with the electron and hole transport materials.
The type and magnitude of the hysteresis can be tuned by bias pre-poling and compositional changes in the PSC structure \cite{doi:10.1002/adfm.201806479}.
However, setting a certain configuration of the PSC, the J-V hysteresis depends on the pre-conditioning and measurement conditions. Conversely, for a well defined measurement protocol, the hysteresis can be correlated with structural properties or changes within the PSCs, the magnitude of ion migration implying potential degradation \cite{doi:10.1021/acs.accounts.5b00420,Kim2017,C6TA09202K}.  

More recently, an increasing number of papers reported hysteresis-free behavior in connection with enhanced performance, in terms of PCE and/or stability. This is typically achieved by improving the electron extraction, while reducing the number of surface traps, either using fullerene derivatives \cite{Xu2015,C5EE00120J,doi:10.1021/acs.jpclett.6b02103}, employing SnO$_2$ as electron transfer layer (ETL) \cite{AENM:AENM201700414,C7TA08040A,doi:10.1002/adfm.201706276} or by optimizing the growth of the perovskite active layer \cite{doi:10.1021/acsenergylett.8b00871,SIDHIK2017169}. In spite of genuine progress, a hysteresis-free behavior does not always necessarily imply steady-state. It is known that employing a large enough scan rate can suppress the hysteretic behavior, although it may not correspond to steady-state. Furthermore, pre-poling the solar cell and performing separate forward and reverse bias scans can, in some cases, render an apparent hysteresis-free behavior.

In order to quantify the hysteresis magnitude and type, a metric was introduced, commonly known as hysteresis index (HI) or hysteresis factor. Although it was widely used in the research community, HI has also been met with reservation \cite{C4EE03664F,Christoforo_2015} and even with harsh criticism \cite{doi:10.1021/acsenergylett.8b01627}, while still being used in recent papers \cite{C8EE01136B,C8EE02404A}. A number of definitions have been proposed, based on the differences between the forward and reverse scans, either using the two currents at given bias, the PCEs or the integrated power output values. These definitions have some shortcommings as pointed out in our recent paper \cite{NEMNES2018976}. However, we argue here that a properly defined HI can be a relevant quantity for at least two reasons. First, by evaluating the hysteresis one may quantify the ion migration and correlate it with aging, which, in turn can give a rapid feedback over the degradation potential. Secondly, one can assess the impact of external fluctuations associated with relatively long timescales on the device operation, e.g. illumination or load variations.

In this paper we discuss the conditions for attaining the hysteresis-free behavior, for both steady-state and non-steady-state conditions, illustrating some common pitfalls in the PCE determination. The measured J-V characteristics are complemented by simulations using the dynamic electrical model (DEM) \cite{NEMNES2017197}. By varying the bias scan rate and the pre-poling conditions, we investigate the hysteresis type and magnitude by defining an HI and indicating its potential use in the context of PSC aging. Furthermore, the analysis of the dynamic regime, not only ensures a correct assessment of the steady state, but also provides insight regarding the characteristic relaxation times corresponding to the slow processes. The interplay between normal and inverted hysteresis is analyzed here in further detail. In addition, switching from normal to inverted hysteresis conditions we identify two different time scales, that are likely to correspond to the migration of two ionic species.

\section{Device fabrication and characterization}

The PSC fabrication starts by depositing onto a FTO coated commercial glass substrate (10 $\Omega$/sq from Xin Yan Technology LTD) of a 100-150 nm compact TiO$_2$ layer by spray pyrolysis at 450 $^\circ$C with a titanium diisopropoxide bis(acetylacetonate) solution (Aldrich) as raw material and nitrogen as a carrier gas. For the deposition of the mesoporous scaffold with a maximum thickness of 350 nm a commercial TiO$_2$ paste (Solaronix Ti-Nanoxide N/SP) was used, a thermal treatment at 500 $^\circ$C for 1h being necessary. The CH$_3$NH$_3$PbI$_{2.6}$Cl$_{0.4}$ halide hybrid perovskite was deposited by a modified one-step method from a 1.415M stoichiometric precursor solution with DMF and DMSO as solvents (7.7:1 weight ratio) to form the active layer. A volume of 100 $\mu$l of diethyl ether was added at second 9 of the spin-coating process (2000 rpm for 25 s), the crystallization of the perovskite being completed by a 100 $^\circ$C, 180 s annealing step, a 250-350 nm perovskite capping layer being formed. Using a solution that contains 80 mg spiro-OMeTAD (Borun Chemical), 28 $\mu$l 4-tert-butylpyridine and 18 $\mu$l of bis(trifluoromethane)sulfonimide lithium salt acetonitrile solution (520 mg/ml) a 250 nm thick spiro-OMeTAD layer was deposited by spin-coating at 1500 rpm for 30 s in a controlled atmosphere (24 $^\circ$C and a humidity of max. 10\%). As a counter electrode, a 100 nm thick gold film was deposited by RF magnetron sputtering, the PSCs having a active area of 0.09 cm$^2$. All layer thicknesses were evaluated using scanning electron microscopy. More details for the fabrication procedure can be found in Ref.\ \cite{doi:10.1021/acs.jpclett.6b02375}.

In addition to this PSC structure, we considered a modified ETL by adding a $\sim$70 nm [6,6]-Phenyl-C61-butyric acid methyl ester (PCBM) layer between the mesoporous TiO$_2$ and the absorber. This layer was fabricated by spin coating through a two step sequence (1800 rpm for 30 s and 2000 rpm for 15 s) using a 20 mg/ml in chlorbenzene solution followed by a short thermal treatment at 100 $^\circ$C for 30 s.
Alternatively, starting from the standard PSC, we varied the absorber composition using a mixed organic cation perovskite, where the methyl-ammonium organic cation was partially substituted by imidazole [(CH$_3$NH$_3$)$_{0.95}$(C$_3$N$_2$H$_4$)$_{0.05}$PbI$_{2.6}$Cl$_{0.4}$], while keeping all the other layers of the standard structure the same.

An Oriel VeroSol-2 Class AAA LED Solar Simulator having AM 1.5G spectrum. The 1 Sun illumination of 100 mW/cm$^2$ (calibrated with a Newport standard silicon solar cell 91150) was performed through a geometrically identical aperture with the Au electrodes, a $3\times3$ mm$^2$.
The J-V characteristics were performed with Keithley 2400 Source Meter, using a bias step of 20 mV. The delay times were varied, while keeping the bias step constant, in order to perform J-V scans at different rates.\\

\begin{figure}[t]
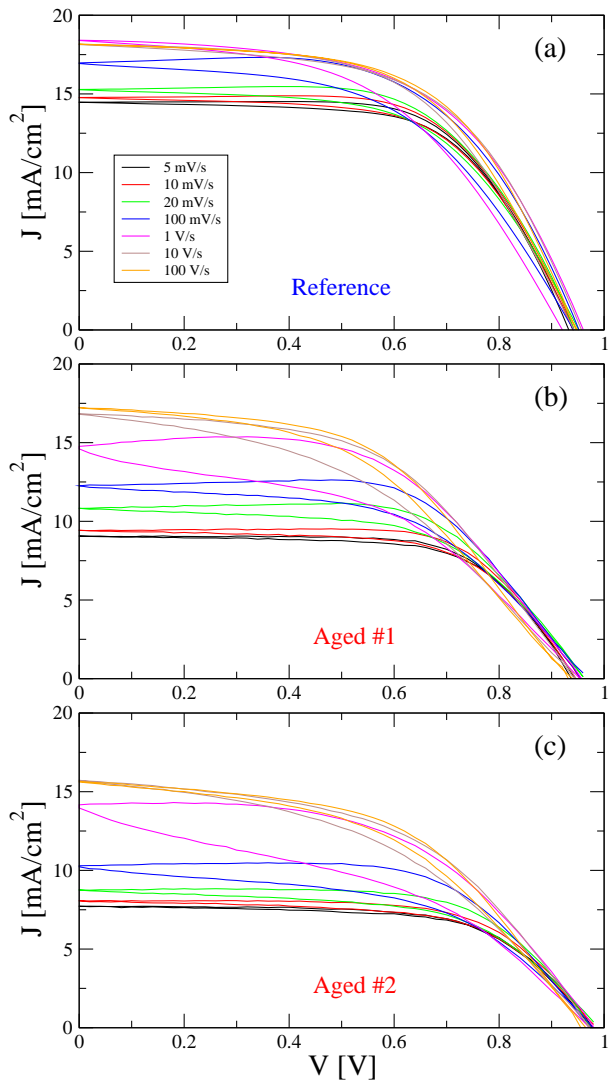

\centering
\includegraphics[width=8.0cm]{figure1a}
\includegraphics[width=8.0cm]{figure1b}
\includegraphics[width=8.0cm]{figure1c}
\caption{Measured J-V characteristics for different scan rates, from 5 mV/s to 100 V/s: (a) reference; (b) sample aged for 1 week; (c) sample aged for 3 weeks. Reverse - forward scans were performed after pre-conditioning at open circuit for 30 s. The hysteresis tends to vanish for very small and very large scan rates.}
\label{IV-alpha}
\end{figure}

\section{Results and discussion}

\begin{figure}[t]
\centering
\includegraphics[width=8.5cm]{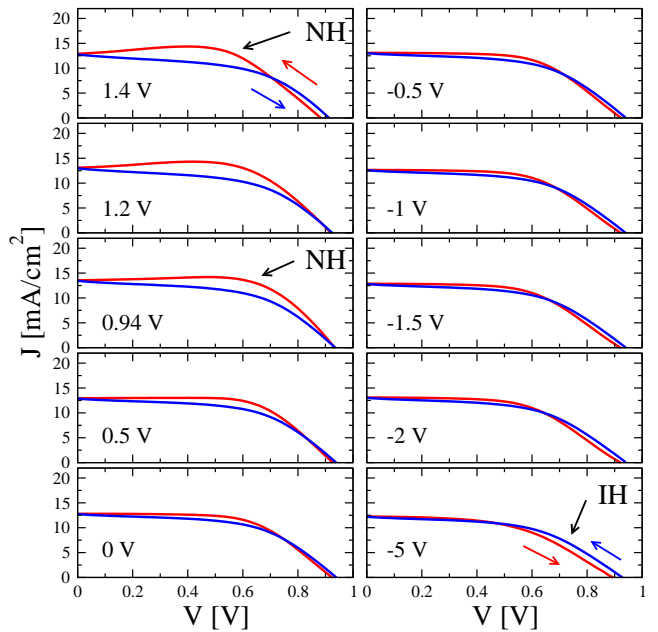}
\caption{Measured J-V characteristics obtained after pre-poling the sample at $V_{\rm pol}$, starting from 1.4 V down to -5 V, for a time $t_{\rm pol} = 30$ s. The reverse - forward scans were performed with a constant scan rate $\alpha = 100$ mV/s. While the hysteresis is tuned from mostly NH to mostly IH behavior, the reverse and forward J-V characteristics can become very similar (e.g. for $V_{\rm pol} = -0.5$ V).  
}
\label{IV-Vpol-100}
\end{figure}

\begin{figure}[t]
\centering
\includegraphics[width=8.5cm]{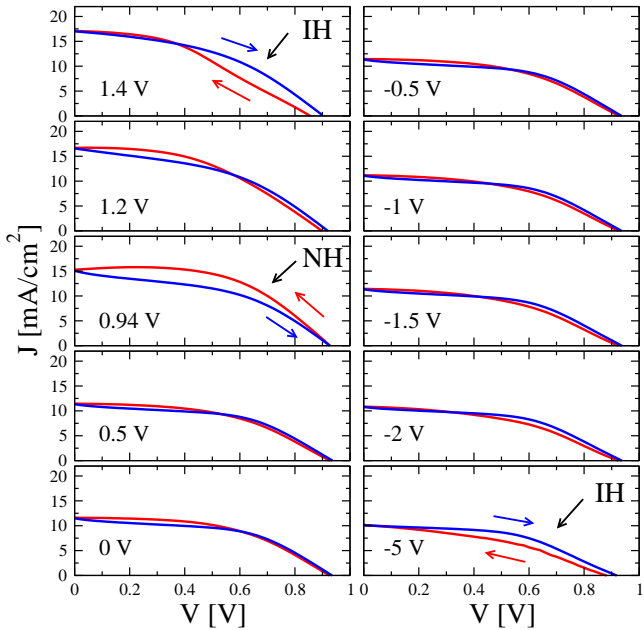}
\caption{Measured J-V characteristics obtained after pre-poling the sample at the same values as in Fig.\ \ref{IV-Vpol-100}, but using a larger scan rate $\alpha = 1000$ mV/s. In this case, IH is found for large positive poling voltages ($V_{\rm pol} = 1.4$ V), in contrast to the smaller scan rate $\alpha = 100$ mV/s, where NH is found.   
}
\label{IV-Vpol-1000}
\end{figure}

\begin{figure}[t]
\centering
\includegraphics[width=8.5cm]{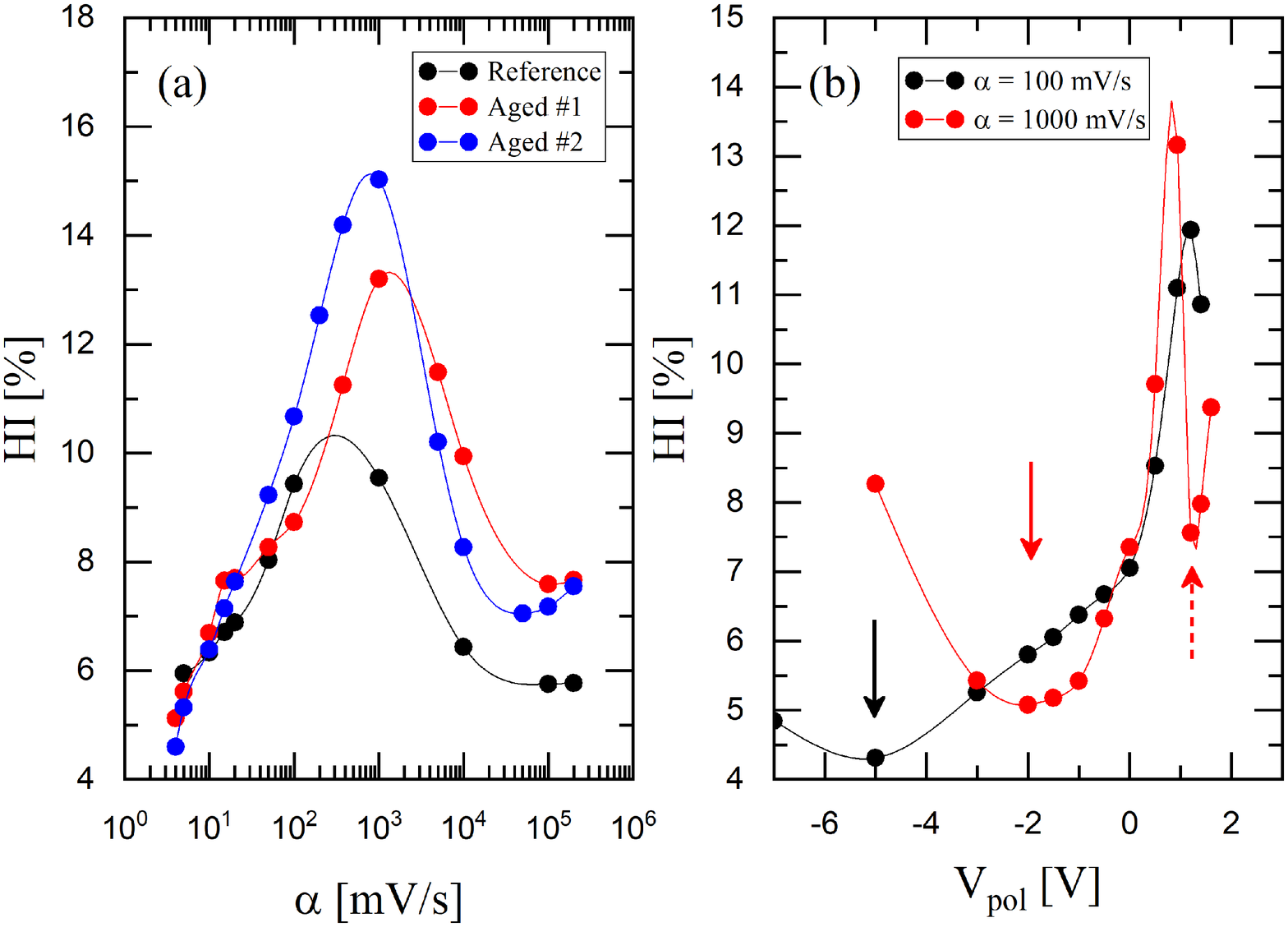}
\caption{The hysteretic behavior observed in Figs.\ \ref{IV-alpha}, \ref{IV-Vpol-100} and \ref{IV-Vpol-1000} is quantified by the hysteresis index (absolute value). (a) The magnitude of the hysteresis increases with aging. (b) An asymmetric behavior is observed for HI for positive and negative poling bias $V_{\rm pol}$. Here, two scan rates, $\alpha = 100$ mV/s (black) and $\alpha = 1000$ mV/s (red), were employed. The arrows (solid lines) mark the minima of the HI's; the dashed arrow marks the mixed hysteresis separating IH and NH for $\alpha = 1000$ mV/s at strong positive poling ($V_{\rm pol} = 1.2$ V). The lines are drawn to guide the eyes.  
}
\label{HI}
\end{figure}

If the steady-state condition is obtained during the forward and the reverse scans, then the J-V characteristics is by definition {\it hysteresis-free}. This condition can be therefore achieved in any sample, provided it does not sustain degradation during the measurement. Thus, the questions {\it how large is the hysteresis?} and {\it what is the type of the hysteresis ?} are only meaningful under clearly specified pre-conditioning and measurement conditions.

A straightforward procedure to obtain the steady-state maximum PCE is to perform a very slow scan, stabilizing each bias point, as pointed out in Ref.\ \cite{doi:10.1021/acsenergylett.8b01627} and discussed in detail in Refs.\ \cite{C4EE02465F,snaith,NEMNES2018976}. In this way, potential ambiguities introduced by dynamic J-V scans with typically faster bias scan rates or maximum power point tracking (MPPT) algorithms are eliminated. However, the slow scan procedure works well only if the J-V characteristics is not affected by degradation during the measurement. Dunbar {\it et al.} provided an extensive inter-laboratory comparison regarding the measurement techniques and discussed the difficulties and limitations introduced by degradation processes relevant to the timescale of the measurement \cite{C7TA05609E}. In this respect, five device classes were outlined, in terms of different combinations of degradation and slow response.

Dynamic J-V scans with moderate scan rates can circumvent the degradation problem and, in addition, can reduce the investigation time. Also, they are frequently employed in the vast majority of studies. However, necessary precautions should be taken in order to obtain a correct evaluation of the steady-state parameters.  This typically requires an analysis over a broad range of bias scan rates and poling conditions, in order to make sure that the estimated result is as close as possible to the correct steady-state value. Only in this way, one can distinguish between the true steady-state hysteresis-free J-V characteristics and other measurement artifacts.

\subsection{Bias scan rates}

We investigate the behavior of the PSCs by varying the bias scan rate $\alpha$ in the range of 5 mV/s -- 100 V/s. Figure\ \ref{IV-alpha}(a) shows measured J-V characteristics, performed as continuous reverse-forward (R-F) scans, the PSCs being pre-poled at $V_{\rm pol} = 0.98$ V for a time $t_{\rm pol} = 30$ s, which, in a good approximation, corresponds to the stabilized open circuit voltage $V_{\rm oc}$. The typical behavior is found: as $\alpha$ increases the hysteresis magnitude has a maximum and the short circuit current is enhanced. The current in the reverse characteristics exhibits an overshoot, which is visible for an intermediate scan rate, when the measurement time interval is of the same order with the relaxation time scale of the slow process. For a slow scan rate of $5$ mV/s an almost hysteresis-free J-V characteristics is obtained, where the steady-state is practically achieved. The maximum PCE is 8.79\% in the reverse scan, found at $V=0.78$ V, and a value of 8.68\% in the forward scan at the same bias. For a fast scan rate 10 V/s again a hysteresis-free behavior is observed. However, in this case the J-V characteristics is far away from the steady-state and the resulting PCE is $\sim$13 \% higher, comparing the reverse characteristics. For the reference sample, the largest hysteresis corresponds to $\alpha \approx 100$ mV/s. Note that within this analysis, continuous reverse-forward scans were employed. If separate reverse and forward scans are performed, the time interval spent between them can introduce uncontrollable depolarization effects. 

The same sample, aged for one week, was further analyzed for the same bias scan rate conditions and the J-V characteristics are shown in Fig.\ \ref{IV-alpha}(b). In this case, compared to the reference sample, the short-circuit current density $J_{sc}$ corresponding practically to the steady-state ($\alpha = 5$ mV/s) is significantly reduced, from 14.5 mA/cm$^2$ to 9.1 mA/cm$^2$. Surprisingly, for $\alpha = 10$ V/s, the hysteresis-free J-V characteristics is quite close to the one obtained for the reference sample, with $J_{sc}$ decreasing from 18.1 mA/cm$^2$ to 17.2 mA/cm$^2$. The hysteresis is maximized at higher scan rates, implying that in the aged sample the relaxation time scale is smaller. The magnitude of the maximal hysteresis will be analyzed later, by evaluating the HI for each measurement set. Investigating again the same sample after 3 weeks with the same procedure, we find the performance parameters at lower values, following the same trend, noticing however a slower rate of the degradation, as shown in Fig.\ \ref{IV-alpha}(c). 

Our results, obtained for a mixed halide (Cl/I) perovskite based PSC in a regular configuration, are consistent with existing data in the literature. The R-F dynamic J-V scans typically show an enhanced short-circuit current as the bias scan rate is increased and a maximum hysteresis at intermediate scan rates \cite{C4EE03664F,doi:10.1021/acs.jpclett.7b00045,C4EE02465F} in standard MAPI based PSCs. These features are also observed independent on the fabrication method of the TiO$_2$ layer, such as spray pyrolysis (current study) vs. spin coating in our earlier studies \cite{NEMNES2017197,doi:10.1021/acs.jpcc.7b04248}. Moreover, both regular and inverted solar cells were shown to exhibit a qualitatively similar behavior, although the magnitude of the hysteretic effects may be different \cite{doi:10.1021/acs.jpcc.6b04233}.
Furthermore, the modified PSCs, obtained by introducing the PCBM layer at the ETL/absorber interface or by considering a mixed composition of organic cations in the perovskite, exhibit a rather similar behavior as the original PSC configuration, as one may see in Fig. S4.
The experimental results are complemented by simulations performed using DEM \cite{NEMNES2017197}, as indicated in Fig.\ S1(a,b,c).  In addition, one should note that a temporary bias-stress induced degradation becomes visible after multiple measurements are performed, as indicated in Fig S2. 

An important conclusion is that, although both the reference and aged samples exhibit relatively small differences in the hysteresis-free J-V characteristics at high scan rates of 10--100 V/s, the steady-state values of the current are quite different. Therefore, employing high scan rates, which has the benefit of avoiding potential degradation during the measurement by reducing the total measurement time, can lead to an erroneous description of the steady-state and, potentially, of the time evolution of the long term degradation, which typically occurs even in passive storage conditions. A more systematic analysis of the degradation is performed by employing the hysteresis index in Section \ref{HI-section}.

\subsection{Bias pre-poling}

In this section we analyze in how far sample poling can render a close to hysteresis-free behavior. Knowing that by tuning the poling voltage, the hysteresis type can be switched from normal to inverted \cite{doi:10.1021/acs.jpcc.7b04248}, we observe the conditions for which the deviation between the two characteristics, reverse and forward, is minimized. In Fig.\ \ref{IV-Vpol-100} a selection of the R-F scans is shown, measured with a scan rate $\alpha = 100$ mV/s, pre-poling at $V_{\rm pol}$ in a range from $-5$ V to $1.4$ V. As detailed in Ref.\ \cite{NEMNES2018976}, the type of the hysteresis depends both on the sign of the poling voltage, but also on the sequence of the two scans, reverse-forward or forward-reverse. For the R-F scans, a positive pre-poling bias induces NH and a negative one render IH, while the opposite is found for F-R scans. 

The inverted hysteresis was first reported by Tress {\it et al.} and was attributed to charge extraction barriers \cite{doi:10.1002/aenm.201600396}. Further on, the IH was shown to be dependent on specific compositions of the perovskite layer, PSC structure, processing or measurement conditions as discussed in a recent review \cite{doi:10.1021/acsenergylett.8b01606}. Regarding our analysis based on poling, it is worth pointing out a notable difference between NH and IH: while a significant NH typically appears for poling voltages just above $V_{\rm oc}$, it is much harder to induce IH, using low negative voltages. This is consistent with  an asymmetry in either migrating species, the charge of the ions/vacancies, or regarding the ion blocking interface (TiO$_2$ or Spiro).  

In Fig.\ \ref{IV-Vpol-100} we observe a quite systematic tuning of the hysteresis, from NH to IH, as the poling voltage decreases. In this sequence, typically a mixed hysteresis is obtained, with a crossing point between forward and reverse scans. This point gradually shifts towards lower voltages, minimizing the hysteresis. It is important to note that applying different poling biases $V_{\rm pol}$ (for a time $t_{\rm pol}$) produces similar effects to setting different starting biases of the scan \cite{NEMNES2017197} in typical measurements. However, using the same poling voltages, but employing a larger scan rate $\alpha = 1000$ mV/s a rather different picture emerges for large positive $V_{\rm pol}$, as indicated in Fig.\ \ref{IV-Vpol-1000}: IH is obtained instead of NH for $V_{\rm pol} = 1.4$ V and a mixed hysteresis follows for $V_{\rm pol} = 1.2$ V, before the NH is recovered for $V_{\rm pol} = 0.94$ V. A similar behavior was observed before by Shen {\it et al.} \cite{doi:10.1021/acs.jpclett.7b00571}, where an IH was reported for fast scan rates (> 1 V/s), while the PSCs showed NH for low scan rates (< 100 mV/s). The reverse J-V characteristics has a typical S-shaped form, when the PSC is pre-poled at large forward bias ($V_{\rm pol} = 1.4$ V). This effect can be explained by an enhanced recombination due to the strong positive poling as reported in Ref.\ \cite{doi:10.1021/acs.jpclett.7b00571}, causing the displacement and accumulation of ions at the interfaces, resulting in a band bending at the perovskite/TiO$_2$ interface, which temporarily impedes electron extraction. Assuming that this process has a much shorter relaxation time (< 1s) compared to the redistribution of ions in the bulk of the perovskite, the strong recombination only affects the first part of the reverse scan when $\alpha \sim 100$ mV/s or smaller, while it is present on the entire R-F scan when $\alpha \sim 1000$ mV/s or larger. This is also consistent with the small inverted hysteresis visible near open circuit in Fig.\ \ref{IV-Vpol-100} for $V_{\rm pol} = 1.4$ V.
The modified PSC structures follow the same trend, as it is indicated in Figs.\ S5 and S6. Introducing the PCBM layer, a NH is found for $V_{\rm pol} = 1.4$ V for $\alpha = 100$ mV/s, while an IH is visible at the larger scan rate $\alpha = 1000$ mV/s, for the same $V_{\rm pol}$, as one can see from Figs.\ S4 and S5. For the samples with mixed cation composition, although a rather small NH is obtained for $V_{\rm pol} = 1.4$ V and  $\alpha = 1000$ mV/s, the NH is further enhanced as $V_{\rm pol}$ is lowered towards $V_{\rm oc}$ (e.g. $V_{\rm pol} =$ 1.2 V, 1.4 V), which confirms the hypothesis of the initial enhanced recombination due to poling.

\subsection{Common pitfalls in dynamic J-V measurements}

As the J-V scans are typically employed in the vast majority of studies for characterizing the performance of the PSCs and the hysteresis phenomena, we discuss some potential pitfalls that should be avoided and propose some recommendations:

1). Bias scan rate:
\begin{itemize}
\item
A large scan rate combined with (unintentional) bias pre-poling can artificially enhance the PCE.   
The J-V characteristics will be far away from the stationary case.
\item        
A large scan rate can misleadingly render a hysteresis-free behavior. Exploring only a narrow interval of bias scan rates can indicate that the hysteresis is either increasing or decreasing in magnitude.
\item
The hysteresis magnitude is maximized for intermediate bias scan rates. Very fast or very slow scan rates will render a vanishing hysteresis. 
\end{itemize}    

2). Bias scan range:
\begin{itemize}
\item
A bias scan range exceeding $(0,V_{\rm oc})$ will introduce additional poling, particularly significant beyond $V_{\rm oc}$, compared the J-V characteristics measured within this interval.
\item
A fixed bias scan range can lead to misinterpretations of dynamical effects when comparing samples with different $V_{\rm oc}$'s. In this context, a degradation affected Voc can pose additional problems. 
\end{itemize}

3). Independent forward (F) and reverse (R) scans:
\begin{itemize}
\item
Independent forward and reverse bias scans can lead to significant errors in evaluating the hysteresis: the initial poling of the individual scans is hard to control; the time between separate F and R scans may lead to uncontrollable depolarization.
\item
Consecutive F-R or R-F scans are recommended, i.e. the second scan is conducted in the opposite sweep direction and begins immediately after the first scan is complete. 
\end{itemize}

\subsection{Hysteresis index}
\label{HI-section}

The usage of a hysteresis index was criticized by several authors. Habisreutinger et al. \cite{doi:10.1021/acsenergylett.8b01627} pointed out that the HI relies on dynamic scans which measure arbitrary non-steady-state values and it does not provide practical information to optimize the steady-state performance. In that paper it is further argued that efficiency should always be used to quantify the performance of PSCs, while any metric that is based on non-steady-state parameters should be avoided. Earlier reports also question the utility of HIs, as it depends on several measurement conditions like scan range and scan rate, as indicated by Tress et al. \cite{C4EE03664F} and Christoforo et al. \cite{Christoforo_2015}.

\begin{figure}[t]
\centering
\includegraphics[width=8.5cm]{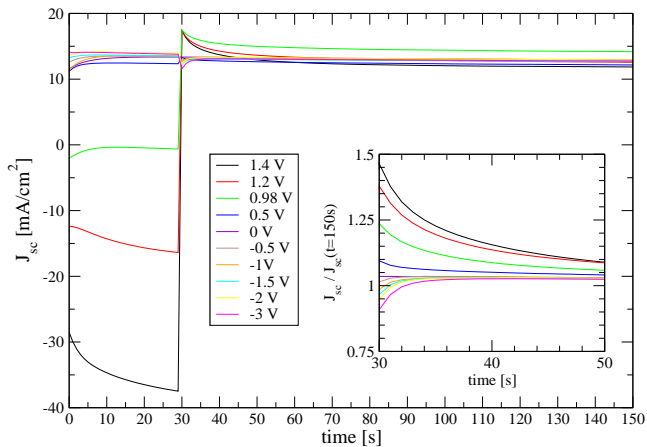}
\caption{The transient current at short-circuit is monitored over time, with the PSC pre-poled at different $V_{\rm pol}$, for a time $t_{\rm pol} = 30$ s. A saturation of the poling condition is observed, yielding a maximal initial $J_{\rm sc}$. Inset: $J_{\rm sc}$ transient currents scaled to the values found at $t = 150$ s. Significantly shorter relaxation timescales are observed for $V_{\rm pol}<0$ compared to $V_{\rm pol}>0$.
}
\label{Isc-t}
\end{figure}

While we recognize some deficiencies of reported HIs, which we aimed to correct in Ref. \cite{NEMNES2018976}, we suggest that systematically performed transient analysis can serve the goal of PSCs optimization, not necessarily directly related to PCE enhancement, but particularly for stability issues such as possibly quantifying ion migration and connecting microscopic parameters such as the relaxation time scale to the solar cell degradation.

Figure\ \ref{HI}(a) shows the behavior of the HI, defined as in Ref.\ \cite{NEMNES2018976}, as a function of the bias scan rate and poling voltage. It was reported by Levine {\it et al.} that HI vs. $\log(\alpha)$ results in a bell-shaped behavior \cite{doi:10.1021/acs.jpcc.6b04233}, which is consistent with our findings. The small values of HI for low and high scan rates correspond to two extreme cases, when the time scale of the measurement is either large enough to exclude the poling effects (for small $\alpha$) or small enough to retain the variation of these additional polarization processes (for large $\alpha$). In between these extreme cases, the measurement time interval becomes comparable with one or more of the relaxation times corresponding to the different possible transient effects. The position of the maximum is therefore related to the relevant relaxation time scale. Interestingly, the aged samples have the bell-shaped distributions of HI's slightly shifted towards larger $\alpha$ values, which would correspond to a faster relaxation. However, a clear trend is visible regarding the magnitude of the hysteresis, which increases steadily as the sample degrades. Another observation is that HI has a larger variation with aging at higher scan rates, in contrast to the lower rates.  Therefore using a narrow range of relatively high scan rates should be avoided both for PCE evaluation and aging analysis. Instead, it is recommendable to explore a wide range of bias scan rates and determine the maximal HI and the two hysteresis-free extremes.   

The bias pre-poling performed at two bias scan rates, $\alpha = 100$ mV/s and $\alpha = 1000$ mV/s, were shown to have qualitatively different behaviors at positive poling voltages (compare Figs.\ \ref{IV-Vpol-100} and \ref{IV-Vpol-1000}). For the large scan rate, IH ($V_{\rm pol} = 1.4$ V) is turned into NH ($V_{\rm pol} = 0.94$ V), while a mixed hysteresis with correspondingly lower magnitude is found for $V_{\rm pol} = 1.2$ V.
Poling at negative bias values produces a rather small IH for both scan rates. However, a potentially significant observation is that the minimum of HI moves towards smaller $|V_{\rm pol}|$, i.e. the IH is enhanced at moderate to low negative biases. This lends the idea that the process generating IH might have a smaller associated relaxation time. For a slow scan rate, if the relaxation time is small enough, IH would take effect only around the open circuit, i.e. at the beginning of the reverse scan. This further analyzed by numerical simulations using DEM, introducing an additional recombination current component with a shorter relaxation time, as indicated in Fig S3. The minima in the HI observed in Fig.\ \ref{HI}(b) reflect a condition of a reduced hysteresis, as it is visible in Figs.\ \ref{IV-Vpol-100} and \ref{IV-Vpol-1000}.

In order to test the assumption that negative poling correlates with shorter relaxation timescales, we monitored the transient short-circuit current ($J_{\rm sc}$), by pre-poling the PSCs at both positive and negative $V_{\rm pol}$, as indicated in Fig.\ \ref{Isc-t}. For $V_{\rm pol}>0$, $J_{\rm sc}$ has an initial value larger than the steady-state value and decays with a timescale ($\sim$ 5 -- 10 s), in contrast to the case of negative poling, where $J_{\rm sc}$ increases towards the steady-state value in a comparatively smaller time ($\sim$ 1 -- 3 s). This indicates that the IH is typically unlikely under negative poling at slow scan rates, but it is enhanced as the scan rate increases, confirming the behavior obtained for $\alpha =100$ mV/s and $\alpha =1000$ mV/s in Fig.\ \ref{HI}(b). The magnitude of the relaxation timescales in correlation with the bias scan range fixes the magnitude and type of the hysteresis, yielding behaviors from heavily hysteretic to hysteresis-free. Therefore, in general, a correct estimation of the relaxation timescales is an essential step in assessing the hysteresis phenomena.

\section{Conclusions}

In any PSC structure, the J-V hysteresis in a relative process, which depends on both pre-conditioning and measurement conditions. Other factors may have a significant contribution to the observed hysteresis, such as temporary or permanent degradation. Here we explored the conditions for obtaining J-V characteristics with minimal hysteresis, frequently referred as hysteresis-free. The bias scan rate and the bias pre-poling voltage are two parameters that can induce apparent hysteresis-free condition in reverse-forward scans, particularly large scan rates and negative poling, where the J-V characteristics may be significantly different from the steady-state. If the scan rate is a controlled parameter in a typical J-V scan, unintentional poling may simply occur due to consecutive measurements or a start bias outside $(0,V_{\rm oc})$ range. Furthermore, to quantify the J-V hysteresis, a properly defined HI is a useful measure. As recognized by other authors although the HI does not address directly the PCE optimization, which is mostly related to the steady-state, however, HI may find its utility in relating consistently the hysteresis magnitude to potential degradation effects. Also, using the HI we pointed out the asymmetry between NH and IH with regard to the poling voltage and the influence of the bias scan rate. 

{Based on these elements we assemble an explicit list of conditions to obtain and properly characterize a hysteresis-free behavior, which may also be subject of evaluation under long term degradation:
\begin{itemize}
\item
Using a slow enough scan rate the steady state may be achieved, which is by definition hysteresis-free, provided the measurement time interval is smaller than the degradation time scale;
\item
A very fast scan renders also a hysteresis-free J-V characteristics, which may be far away from the steady state, as the slow process (ion migration) cannot follow;
\item
A suitable bias pre-poling, which may be accidentally induced by consecutive measurements, can reduce the apparent hysteresis (e.g. $V_{\rm pol}<0$ in R-F scans);
\item
The {\it maximal hysteresis} should to be determined for a set of relevant preconditioning and measurement conditions (e.g. by varying the bias scan rate and poling voltage) and should be smaller than a specified threshold value in order for the PSC to qualify for the status of {\it hysteresis-free};
\item
A hysteresis-free behavior under certain bias pre-poling (typically $V_{\rm pol}<0$) may indicate a diminished or even the absence of a combined process, consisting of ion migration and blocking effect at the respective interface (e.g. our data suggests that the migration of I$^-$ to TiO$_2$ or MA$^+$ to Spiro in the case of $V_{\rm pol}<0$ is less likely than the opposite process, i.e. the migration of  MA$^+$ to TiO$_2$ or I$^-$ to Spiro, for $V_{\rm pol}>0$), which can be revealed by the relaxation time scales of current transients;
\item
The bell-shaped plots of the hysteresis index vs. the logarithm of the scan rate provides a connection between the magnitude of the hysteretic effects and sample aging, which is particularly important as a reduced hysteresis may signal a diminished ion migration.
\end{itemize}}

\begin{acknowledgments}
This work was supported by the National Ministry of Research and Innovation under the projects PN18-090205, PN19-030101 and by UEFISCDI under grant PN-III-P1-1.1-PD-2016-1546 and PN-III-P4-ID-PCE-2016-0692.
\end{acknowledgments}

\bibliography{manuscript_R1} 

\providecommand*{\mcitethebibliography}{\thebibliography}
\csname @ifundefined\endcsname{endmcitethebibliography}
{\let\endmcitethebibliography\endthebibliography}{}
\begin{mcitethebibliography}{39}
\providecommand*{\natexlab}[1]{#1}
\providecommand*{\mciteSetBstSublistMode}[1]{}
\providecommand*{\mciteSetBstMaxWidthForm}[2]{}
\providecommand*{\mciteBstWouldAddEndPuncttrue}
  {\def\EndOfBibitem{\unskip.}}
\providecommand*{\mciteBstWouldAddEndPunctfalse}
  {\let\EndOfBibitem\relax}
\providecommand*{\mciteSetBstMidEndSepPunct}[3]{}
\providecommand*{\mciteSetBstSublistLabelBeginEnd}[3]{}
\providecommand*{\EndOfBibitem}{}
\mciteSetBstSublistMode{f}
\mciteSetBstMaxWidthForm{subitem}
{(\emph{\alph{mcitesubitemcount}})}
\mciteSetBstSublistLabelBeginEnd{\mcitemaxwidthsubitemform\space}
{\relax}{\relax}

\bibitem[nre()]{nrel}
\emph{Best Research Cells Efficiencies from NREL (2018),
  https://www.nrel.gov/pv/assets/pdfs/pv-efficiencies-07-17-2018.pdf}\relax
\mciteBstWouldAddEndPuncttrue
\mciteSetBstMidEndSepPunct{\mcitedefaultmidpunct}
{\mcitedefaultendpunct}{\mcitedefaultseppunct}\relax
\EndOfBibitem
\bibitem[Besleaga \emph{et~al.}(2016)Besleaga, Abramiuc, Stancu, Tomulescu,
  Sima, Trinca, Plugaru, Pintilie, Nemnes, Iliescu, Svavarsson, Manolescu, and
  Pintilie]{doi:10.1021/acs.jpclett.6b02375}
C.~Besleaga, L.~E. Abramiuc, V.~Stancu, A.~G. Tomulescu, M.~Sima, L.~Trinca,
  N.~Plugaru, L.~Pintilie, G.~A. Nemnes, M.~Iliescu, H.~G. Svavarsson,
  A.~Manolescu and I.~Pintilie, \emph{J. Phys. Chem. Lett.}, 2016, \textbf{7},
  5168--5175\relax
\mciteBstWouldAddEndPuncttrue
\mciteSetBstMidEndSepPunct{\mcitedefaultmidpunct}
{\mcitedefaultendpunct}{\mcitedefaultseppunct}\relax
\EndOfBibitem
\bibitem[Xiong \emph{et~al.}(2016)Xiong, Yang, Cao, Wu, Huang, Sun, Zhang, Liu,
  Tao, Gao, and Yang]{XIONG201630}
J.~Xiong, B.~Yang, C.~Cao, R.~Wu, Y.~Huang, J.~Sun, J.~Zhang, C.~Liu, S.~Tao,
  Y.~Gao and J.~Yang, \emph{Organic Electronics}, 2016, \textbf{30}, 30 --
  35\relax
\mciteBstWouldAddEndPuncttrue
\mciteSetBstMidEndSepPunct{\mcitedefaultmidpunct}
{\mcitedefaultendpunct}{\mcitedefaultseppunct}\relax
\EndOfBibitem
\bibitem[Ginting \emph{et~al.}(2017)Ginting, Jeon, Lee, Jin, Kim, and
  Kang]{C6TA09202K}
R.~T. Ginting, M.-K. Jeon, K.-J. Lee, W.-Y. Jin, T.-W. Kim and J.-W. Kang,
  \emph{J. Mater. Chem. A}, 2017, \textbf{5}, 4527--4534\relax
\mciteBstWouldAddEndPuncttrue
\mciteSetBstMidEndSepPunct{\mcitedefaultmidpunct}
{\mcitedefaultendpunct}{\mcitedefaultseppunct}\relax
\EndOfBibitem
\bibitem[Dao \emph{et~al.}(2017)Dao, Tsuji, Fujii, and Ozaki]{DAO2017229}
Q.-D. Dao, R.~Tsuji, A.~Fujii and M.~Ozaki, \emph{Organic Electronics}, 2017,
  \textbf{43}, 229 -- 234\relax
\mciteBstWouldAddEndPuncttrue
\mciteSetBstMidEndSepPunct{\mcitedefaultmidpunct}
{\mcitedefaultendpunct}{\mcitedefaultseppunct}\relax
\EndOfBibitem
\bibitem[Tress(2017)]{doi:10.1021/acs.jpclett.7b00975}
W.~Tress, \emph{The Journal of Physical Chemistry Letters}, 2017, \textbf{8},
  3106--3114\relax
\mciteBstWouldAddEndPuncttrue
\mciteSetBstMidEndSepPunct{\mcitedefaultmidpunct}
{\mcitedefaultendpunct}{\mcitedefaultseppunct}\relax
\EndOfBibitem
\bibitem[Juarez-Perez \emph{et~al.}(2014)Juarez-Perez, Sanchez, Badia,
  Garcia-Belmonte, Kang, Mora-Sero, and Bisquert]{doi:10.1021/jz5011169}
E.~J. Juarez-Perez, R.~S. Sanchez, L.~Badia, G.~Garcia-Belmonte, Y.~S. Kang,
  I.~Mora-Sero and J.~Bisquert, \emph{The Journal of Physical Chemistry
  Letters}, 2014, \textbf{5}, 2390--2394\relax
\mciteBstWouldAddEndPuncttrue
\mciteSetBstMidEndSepPunct{\mcitedefaultmidpunct}
{\mcitedefaultendpunct}{\mcitedefaultseppunct}\relax
\EndOfBibitem
\bibitem[Wei \emph{et~al.}(2014)Wei, Zhao, Li, Li, Pan, Xu, Zhao, and
  Yu]{doi:10.1021/jz502111u}
J.~Wei, Y.~Zhao, H.~Li, G.~Li, J.~Pan, D.~Xu, Q.~Zhao and D.~Yu, \emph{The
  Journal of Physical Chemistry Letters}, 2014, \textbf{5}, 3937--3945\relax
\mciteBstWouldAddEndPuncttrue
\mciteSetBstMidEndSepPunct{\mcitedefaultmidpunct}
{\mcitedefaultendpunct}{\mcitedefaultseppunct}\relax
\EndOfBibitem
\bibitem[Chen \emph{et~al.}(2015)Chen, Sakai, Ikegami, and
  Miyasaka]{doi:10.1021/jz502429u}
H.-W. Chen, N.~Sakai, M.~Ikegami and T.~Miyasaka, \emph{The Journal of Physical
  Chemistry Letters}, 2015, \textbf{6}, 164--169\relax
\mciteBstWouldAddEndPuncttrue
\mciteSetBstMidEndSepPunct{\mcitedefaultmidpunct}
{\mcitedefaultendpunct}{\mcitedefaultseppunct}\relax
\EndOfBibitem
\bibitem[Frost \emph{et~al.}(2014)Frost, Butler, and
  Walsh]{doi:10.1063/1.4890246}
J.~M. Frost, K.~T. Butler and A.~Walsh, \emph{APL Materials}, 2014, \textbf{2},
  081506\relax
\mciteBstWouldAddEndPuncttrue
\mciteSetBstMidEndSepPunct{\mcitedefaultmidpunct}
{\mcitedefaultendpunct}{\mcitedefaultseppunct}\relax
\EndOfBibitem
\bibitem[van Reenen \emph{et~al.}(2015)van Reenen, Kemerink, and
  Snaith]{doi:10.1021/acs.jpclett.5b01645}
S.~van Reenen, M.~Kemerink and H.~J. Snaith, \emph{The Journal of Physical
  Chemistry Letters}, 2015, \textbf{6}, 3808--3814\relax
\mciteBstWouldAddEndPuncttrue
\mciteSetBstMidEndSepPunct{\mcitedefaultmidpunct}
{\mcitedefaultendpunct}{\mcitedefaultseppunct}\relax
\EndOfBibitem
\bibitem[Ravishankar \emph{et~al.}(2017)Ravishankar, Almora,
  Echeverria-Arrondo, Ghahremanirad, Aranda, Guerrero, Fabregat-Santiago,
  Zaban, Garcia-Belmonte, and Bisquert]{doi:10.1021/acs.jpclett.7b00045}
S.~Ravishankar, O.~Almora, C.~Echeverria-Arrondo, E.~Ghahremanirad, C.~Aranda,
  A.~Guerrero, F.~Fabregat-Santiago, A.~Zaban, G.~Garcia-Belmonte and
  J.~Bisquert, \emph{The Journal of Physical Chemistry Letters}, 2017,
  \textbf{8}, 915--921\relax
\mciteBstWouldAddEndPuncttrue
\mciteSetBstMidEndSepPunct{\mcitedefaultmidpunct}
{\mcitedefaultendpunct}{\mcitedefaultseppunct}\relax
\EndOfBibitem
\bibitem[Tress \emph{et~al.}(2015)Tress, Marinova, Moehl, Zakeeruddin,
  Nazeeruddin, and Gratzel]{C4EE03664F}
W.~Tress, N.~Marinova, T.~Moehl, S.~M. Zakeeruddin, M.~K. Nazeeruddin and
  M.~Gratzel, \emph{Energy Environ. Sci.}, 2015, \textbf{8}, 995--1004\relax
\mciteBstWouldAddEndPuncttrue
\mciteSetBstMidEndSepPunct{\mcitedefaultmidpunct}
{\mcitedefaultendpunct}{\mcitedefaultseppunct}\relax
\EndOfBibitem
\bibitem[Lopez-Varo \emph{et~al.}(2018)Lopez-Varo, Jimenez-Tejada,
  Garcia-Rosell, Ravishankar, Garcia-Belmonte, Bisquert, and
  Almora]{doi:10.1002/aenm.201702772}
P.~Lopez-Varo, J.~A. Jimenez-Tejada, M.~Garcia-Rosell, S.~Ravishankar,
  G.~Garcia-Belmonte, J.~Bisquert and O.~Almora, \emph{Advanced Energy
  Materials}, 2018, \textbf{8}, 1702772\relax
\mciteBstWouldAddEndPuncttrue
\mciteSetBstMidEndSepPunct{\mcitedefaultmidpunct}
{\mcitedefaultendpunct}{\mcitedefaultseppunct}\relax
\EndOfBibitem
\bibitem[Pham \emph{et~al.}(2018)Pham, Zhang, Tiong, Zhang, Will, Bou,
  Bisquert, Shaw, Du, Wilson, and Wang]{doi:10.1002/adfm.201806479}
N.~D. Pham, C.~Zhang, V.~T. Tiong, S.~Zhang, G.~Will, A.~Bou, J.~Bisquert,
  P.~E. Shaw, A.~Du, G.~J. Wilson and H.~Wang, \emph{Advanced Functional
  Materials}, 2018, \textbf{0}, 1806479\relax
\mciteBstWouldAddEndPuncttrue
\mciteSetBstMidEndSepPunct{\mcitedefaultmidpunct}
{\mcitedefaultendpunct}{\mcitedefaultseppunct}\relax
\EndOfBibitem
\bibitem[Yuan and Huang(2016)]{doi:10.1021/acs.accounts.5b00420}
Y.~Yuan and J.~Huang, \emph{Accounts of Chemical Research}, 2016, \textbf{49},
  286--293\relax
\mciteBstWouldAddEndPuncttrue
\mciteSetBstMidEndSepPunct{\mcitedefaultmidpunct}
{\mcitedefaultendpunct}{\mcitedefaultseppunct}\relax
\EndOfBibitem
\bibitem[Kim \emph{et~al.}(2017)Kim, Bae, Lee, Cho, Lee, Kim, Park, Kwon, Ahn,
  Lee, Kang, Lee, and Kim]{Kim2017}
S.~Kim, S.~Bae, S.-W. Lee, K.~Cho, K.~D. Lee, H.~Kim, S.~Park, G.~Kwon, S.-W.
  Ahn, H.-M. Lee, Y.~Kang, H.-S. Lee and D.~Kim, \emph{Scientific Reports},
  2017, \textbf{7}, 1200\relax
\mciteBstWouldAddEndPuncttrue
\mciteSetBstMidEndSepPunct{\mcitedefaultmidpunct}
{\mcitedefaultendpunct}{\mcitedefaultseppunct}\relax
\EndOfBibitem
\bibitem[Xu \emph{et~al.}(2015)Xu, Buin, Ip, Li, Voznyy, Comin, Yuan, Jeon,
  Ning, McDowell, Kanjanaboos, Sun, Lan, Quan, Kim, Hill, Maksymovych, and
  Sargent]{Xu2015}
J.~Xu, A.~Buin, A.~H. Ip, W.~Li, O.~Voznyy, R.~Comin, M.~Yuan, S.~Jeon,
  Z.~Ning, J.~J. McDowell, P.~Kanjanaboos, J.-P. Sun, X.~Lan, L.~N. Quan, D.~H.
  Kim, I.~G. Hill, P.~Maksymovych and E.~H. Sargent, \emph{Nat. Commun.}, 2015,
  \textbf{6}, 7081\relax
\mciteBstWouldAddEndPuncttrue
\mciteSetBstMidEndSepPunct{\mcitedefaultmidpunct}
{\mcitedefaultendpunct}{\mcitedefaultseppunct}\relax
\EndOfBibitem
\bibitem[Heo \emph{et~al.}(2015)Heo, Han, Kim, Ahn, and Im]{C5EE00120J}
J.~H. Heo, H.~J. Han, D.~Kim, T.~K. Ahn and S.~H. Im, \emph{Energy Environ.
  Sci.}, 2015, \textbf{8}, 1602--1608\relax
\mciteBstWouldAddEndPuncttrue
\mciteSetBstMidEndSepPunct{\mcitedefaultmidpunct}
{\mcitedefaultendpunct}{\mcitedefaultseppunct}\relax
\EndOfBibitem
\bibitem[Valles-Pelarda \emph{et~al.}(2016)Valles-Pelarda, Hames,
  Garcia-Benito, Almora, Molina-Ontoria, Sanchez, Garcia-Belmonte, Martin, and
  Mora-Sero]{doi:10.1021/acs.jpclett.6b02103}
M.~Valles-Pelarda, B.~C. Hames, I.~Garcia-Benito, O.~Almora, A.~Molina-Ontoria,
  R.~S. Sanchez, G.~Garcia-Belmonte, N.~Martin and I.~Mora-Sero, \emph{J. Phys.
  Chem. Lett.}, 2016, \textbf{7}, 4622--4628\relax
\mciteBstWouldAddEndPuncttrue
\mciteSetBstMidEndSepPunct{\mcitedefaultmidpunct}
{\mcitedefaultendpunct}{\mcitedefaultseppunct}\relax
\EndOfBibitem
\bibitem[Wang \emph{et~al.}(2017)Wang, Xiao, Yu, Zhao, Awni, Grice, Ghimire,
  Constantinou, Liao, Cimaroli, Liu, Chen, Podraza, Jiang, Al-Jassim, Zhao, and
  Yan]{AENM:AENM201700414}
C.~Wang, C.~Xiao, Y.~Yu, D.~Zhao, R.~A. Awni, C.~R. Grice, K.~Ghimire,
  I.~Constantinou, W.~Liao, A.~J. Cimaroli, P.~Liu, J.~Chen, N.~J. Podraza,
  C.-S. Jiang, M.~M. Al-Jassim, X.~Zhao and Y.~Yan, \emph{Adv. Energy Mater.},
  2017, \textbf{7}, 1700414\relax
\mciteBstWouldAddEndPuncttrue
\mciteSetBstMidEndSepPunct{\mcitedefaultmidpunct}
{\mcitedefaultendpunct}{\mcitedefaultseppunct}\relax
\EndOfBibitem
\bibitem[Jung \emph{et~al.}(2017)Jung, Seo, Lee, Shin, and Park]{C7TA08040A}
K.-H. Jung, J.-Y. Seo, S.~Lee, H.~Shin and N.-G. Park, \emph{J. Mater. Chem.
  A}, 2017, \textbf{5}, 24790--24803\relax
\mciteBstWouldAddEndPuncttrue
\mciteSetBstMidEndSepPunct{\mcitedefaultmidpunct}
{\mcitedefaultendpunct}{\mcitedefaultseppunct}\relax
\EndOfBibitem
\bibitem[Liangbin \emph{et~al.}(2018)Liangbin, Minchao, Cong, Jian, Guang,
  Yaxiong, Junjie, Qi, Pingli, Songzhan, and
  Guojia]{doi:10.1002/adfm.201706276}
X.~Liangbin, Q.~Minchao, C.~Cong, W.~Jian, Y.~Guang, G.~Yaxiong, M.~Junjie,
  Z.~Qi, Q.~Pingli, L.~Songzhan and F.~Guojia, \emph{Advanced Functional
  Materials}, 2018, \textbf{28}, 1706276\relax
\mciteBstWouldAddEndPuncttrue
\mciteSetBstMidEndSepPunct{\mcitedefaultmidpunct}
{\mcitedefaultendpunct}{\mcitedefaultseppunct}\relax
\EndOfBibitem
\bibitem[Pantaler \emph{et~al.}(2018)Pantaler, Cho, Queloz, Garcia~Benito,
  Fettkenhauer, Anusca, Nazeeruddin, Lupascu, and
  Grancini]{doi:10.1021/acsenergylett.8b00871}
M.~Pantaler, K.~T. Cho, V.~I.~E. Queloz, I.~Garcia~Benito, C.~Fettkenhauer,
  I.~Anusca, M.~K. Nazeeruddin, D.~C. Lupascu and G.~Grancini, \emph{ACS Energy
  Letters}, 2018, \textbf{3}, 1781--1786\relax
\mciteBstWouldAddEndPuncttrue
\mciteSetBstMidEndSepPunct{\mcitedefaultmidpunct}
{\mcitedefaultendpunct}{\mcitedefaultseppunct}\relax
\EndOfBibitem
\bibitem[Sidhik \emph{et~al.}(2017)Sidhik, Esparza, Martinez-Benitez,
  Lopez-Luke, Carriles, and la~Rosa]{SIDHIK2017169}
S.~Sidhik, D.~Esparza, A.~Martinez-Benitez, T.~Lopez-Luke, R.~Carriles and
  E.~D. la~Rosa, \emph{Journal of Power Sources}, 2017, \textbf{365}, 169 --
  178\relax
\mciteBstWouldAddEndPuncttrue
\mciteSetBstMidEndSepPunct{\mcitedefaultmidpunct}
{\mcitedefaultendpunct}{\mcitedefaultseppunct}\relax
\EndOfBibitem
\bibitem[Christoforo \emph{et~al.}(2015)Christoforo, Hoke, McGehee, and
  Unger]{Christoforo_2015}
M.~Christoforo, E.~Hoke, M.~McGehee and E.~Unger, \emph{Photonics}, 2015,
  \textbf{2}, 1101--1115\relax
\mciteBstWouldAddEndPuncttrue
\mciteSetBstMidEndSepPunct{\mcitedefaultmidpunct}
{\mcitedefaultendpunct}{\mcitedefaultseppunct}\relax
\EndOfBibitem
\bibitem[Habisreutinger \emph{et~al.}(2018)Habisreutinger, Noel, and
  Snaith]{doi:10.1021/acsenergylett.8b01627}
S.~N. Habisreutinger, N.~K. Noel and H.~J. Snaith, \emph{ACS Energy Letters},
  2018, \textbf{3}, 2472--2476\relax
\mciteBstWouldAddEndPuncttrue
\mciteSetBstMidEndSepPunct{\mcitedefaultmidpunct}
{\mcitedefaultendpunct}{\mcitedefaultseppunct}\relax
\EndOfBibitem
\bibitem[Fassl \emph{et~al.}(2018)Fassl, Lami, Bausch, Wang, Klug, Snaith, and
  Vaynzof]{C8EE01136B}
P.~Fassl, V.~Lami, A.~Bausch, Z.~Wang, M.~T. Klug, H.~J. Snaith and Y.~Vaynzof,
  \emph{Energy Environ. Sci.}, 2018, \textbf{-}, doi:10.1039/C8EE01136B\relax
\mciteBstWouldAddEndPuncttrue
\mciteSetBstMidEndSepPunct{\mcitedefaultmidpunct}
{\mcitedefaultendpunct}{\mcitedefaultseppunct}\relax
\EndOfBibitem
\bibitem[Tavakoli \emph{et~al.}(2018)Tavakoli, Tress, Milic, Kubicki, Emsley,
  and Gratzel]{C8EE02404A}
M.~M. Tavakoli, W.~Tress, J.~Milic, D.~Kubicki, L.~Emsley and M.~Gratzel,
  \emph{Energy Environ. Sci.}, 2018,  3310--3320\relax
\mciteBstWouldAddEndPuncttrue
\mciteSetBstMidEndSepPunct{\mcitedefaultmidpunct}
{\mcitedefaultendpunct}{\mcitedefaultseppunct}\relax
\EndOfBibitem
\bibitem[Nemnes \emph{et~al.}(2018)Nemnes, Besleaga, Tomulescu, Palici,
  Pintilie, Manolescu, and Pintilie]{NEMNES2018976}
G.~A. Nemnes, C.~Besleaga, A.~G. Tomulescu, A.~Palici, L.~Pintilie,
  A.~Manolescu and I.~Pintilie, \emph{Solar Energy}, 2018, \textbf{173}, 976 --
  983\relax
\mciteBstWouldAddEndPuncttrue
\mciteSetBstMidEndSepPunct{\mcitedefaultmidpunct}
{\mcitedefaultendpunct}{\mcitedefaultseppunct}\relax
\EndOfBibitem
\bibitem[Nemnes \emph{et~al.}(2017)Nemnes, Besleaga, Tomulescu, Pintilie,
  Pintilie, Torfason, and Manolescu]{NEMNES2017197}
G.~A. Nemnes, C.~Besleaga, A.~G. Tomulescu, I.~Pintilie, L.~Pintilie,
  K.~Torfason and A.~Manolescu, \emph{Sol. Energy Mater. Sol. Cells}, 2017,
  \textbf{159}, 197 -- 203\relax
\mciteBstWouldAddEndPuncttrue
\mciteSetBstMidEndSepPunct{\mcitedefaultmidpunct}
{\mcitedefaultendpunct}{\mcitedefaultseppunct}\relax
\EndOfBibitem
\bibitem[Unger \emph{et~al.}(2014)Unger, Hoke, Bailie, Nguyen, Bowring,
  Heumuller, Christoforo, and McGehee]{C4EE02465F}
E.~L. Unger, E.~T. Hoke, C.~D. Bailie, W.~H. Nguyen, A.~R. Bowring,
  T.~Heumuller, M.~G. Christoforo and M.~D. McGehee, \emph{Energy Environ.
  Sci.}, 2014, \textbf{7}, 3690--3698\relax
\mciteBstWouldAddEndPuncttrue
\mciteSetBstMidEndSepPunct{\mcitedefaultmidpunct}
{\mcitedefaultendpunct}{\mcitedefaultseppunct}\relax
\EndOfBibitem
\bibitem[Snaith \emph{et~al.}(2014)Snaith, Abate, Ball, Eperon, Leijtens, Noel,
  Stranks, Wang, Wojciechowski, and Zhang]{snaith}
H.~J. Snaith, A.~Abate, J.~M. Ball, G.~E. Eperon, T.~Leijtens, N.~K. Noel,
  S.~D. Stranks, J.~T.-W. Wang, K.~Wojciechowski and W.~Zhang, \emph{J. Phys.
  Chem. Lett.}, 2014, \textbf{5}, 1511--1515\relax
\mciteBstWouldAddEndPuncttrue
\mciteSetBstMidEndSepPunct{\mcitedefaultmidpunct}
{\mcitedefaultendpunct}{\mcitedefaultseppunct}\relax
\EndOfBibitem
\bibitem[Dunbar \emph{et~al.}(2017)Dunbar, Duck, Moriarty, Anderson, Duffy,
  Fell, Kim, Ho-Baillie, Vak, Duong, Wu, Weber, Pascoe, Cheng, Lin, Burn,
  Bhattacharjee, Wang, and Wilson]{C7TA05609E}
R.~B. Dunbar, B.~C. Duck, T.~Moriarty, K.~F. Anderson, N.~Duffy, C.~J. Fell,
  J.~Kim, A.~Ho-Baillie, D.~Vak, T.~Duong, Y.~Wu, K.~Weber, A.~Pascoe, Y.-B.
  Cheng, Q.~Lin, P.~L. Burn, R.~Bhattacharjee, H.~Wang and G.~J. Wilson,
  \emph{J. Mater. Chem. A}, 2017, \textbf{5}, 22542--22558\relax
\mciteBstWouldAddEndPuncttrue
\mciteSetBstMidEndSepPunct{\mcitedefaultmidpunct}
{\mcitedefaultendpunct}{\mcitedefaultseppunct}\relax
\EndOfBibitem
\bibitem[Nemnes \emph{et~al.}(2017)Nemnes, Besleaga, Stancu, Dogaru, Leonat,
  Pintilie, Torfason, Ilkov, Manolescu, and
  Pintilie]{doi:10.1021/acs.jpcc.7b04248}
G.~A. Nemnes, C.~Besleaga, V.~Stancu, D.~E. Dogaru, L.~N. Leonat, L.~Pintilie,
  K.~Torfason, M.~Ilkov, A.~Manolescu and I.~Pintilie, \emph{J. Phys. Chem. C},
  2017, \textbf{121}, 11207--11214\relax
\mciteBstWouldAddEndPuncttrue
\mciteSetBstMidEndSepPunct{\mcitedefaultmidpunct}
{\mcitedefaultendpunct}{\mcitedefaultseppunct}\relax
\EndOfBibitem
\bibitem[Levine \emph{et~al.}(2016)Levine, Nayak, Wang, Sakai, Van~Reenen,
  Brenner, Mukhopadhyay, Snaith, Hodes, and
  Cahen]{doi:10.1021/acs.jpcc.6b04233}
I.~Levine, P.~K. Nayak, J.~T.-W. Wang, N.~Sakai, S.~Van~Reenen, T.~M. Brenner,
  S.~Mukhopadhyay, H.~J. Snaith, G.~Hodes and D.~Cahen, \emph{The Journal of
  Physical Chemistry C}, 2016, \textbf{120}, 16399--16411\relax
\mciteBstWouldAddEndPuncttrue
\mciteSetBstMidEndSepPunct{\mcitedefaultmidpunct}
{\mcitedefaultendpunct}{\mcitedefaultseppunct}\relax
\EndOfBibitem
\bibitem[Tress \emph{et~al.}(2016)Tress, Correa~Baena, Saliba, Abate, and
  Graetzel]{doi:10.1002/aenm.201600396}
W.~Tress, J.~P. Correa~Baena, M.~Saliba, A.~Abate and M.~Graetzel,
  \emph{Advanced Energy Materials}, 2016, \textbf{6}, 1600396\relax
\mciteBstWouldAddEndPuncttrue
\mciteSetBstMidEndSepPunct{\mcitedefaultmidpunct}
{\mcitedefaultendpunct}{\mcitedefaultseppunct}\relax
\EndOfBibitem
\bibitem[Wu \emph{et~al.}(2018)Wu, Pathak, Chen, Wang, Bahrami, Zhang, and
  Qiao]{doi:10.1021/acsenergylett.8b01606}
F.~Wu, R.~Pathak, K.~Chen, G.~Wang, B.~Bahrami, W.-H. Zhang and Q.~Qiao,
  \emph{ACS Energy Letters}, 2018, \textbf{3}, 2457--2460\relax
\mciteBstWouldAddEndPuncttrue
\mciteSetBstMidEndSepPunct{\mcitedefaultmidpunct}
{\mcitedefaultendpunct}{\mcitedefaultseppunct}\relax
\EndOfBibitem
\bibitem[Shen \emph{et~al.}(2017)Shen, Jacobs, Wu, Duong, Peng, Wen, Fu,
  Karuturi, White, Weber, and Catchpole]{doi:10.1021/acs.jpclett.7b00571}
H.~Shen, D.~A. Jacobs, Y.~Wu, T.~Duong, J.~Peng, X.~Wen, X.~Fu, S.~K. Karuturi,
  T.~P. White, K.~Weber and K.~R. Catchpole, \emph{The Journal of Physical
  Chemistry Letters}, 2017, \textbf{8}, 2672--2680\relax
\mciteBstWouldAddEndPuncttrue
\mciteSetBstMidEndSepPunct{\mcitedefaultmidpunct}
{\mcitedefaultendpunct}{\mcitedefaultseppunct}\relax
\EndOfBibitem
\end{mcitethebibliography}
\bibliographystyle{rsc} 

\newpage
\onecolumngrid

\appendix*
\section{Supplementary Information}

\renewcommand{\thefigure}{S\arabic{figure}}
\setcounter{figure}{0}

\subsection{Simulated dynamic J-V characteristics}

Simulations corresponding to the reference and aged samples are performed using DEM. For the reference PSC, the equivalent circuit model parameters are: the series resistance $R_{\rm s} = 120$ $\Omega$, the shunt resistance $R_{\rm sh} = 2$ k$\Omega$, the ideality factor $n = 1.55$, the diode saturation current $J_{\rm s} =1.1$ pA/cm$^2$, the photogenerated current $J_{\rm ph}=14.4$ mA/cm$^2$ and steady-state polarization at open circuit $P_\infty = 12$ mC/cm$^2$. As suggested by the experimental data in Fig.\ 1, aged samples differ significantly by the photogenerated currents, $J_{\rm ph}=8.9$ mA/cm$^2$ (after 1 week) and $J_{\rm ph}=7.4$ mA/cm$^2$ (after 3 weeks). Also, the series resistance increases after one week to $200$ $\Omega$ for the aged PSCs and more ions are accumulating at the interface as the initial polarization increases.

\begin{figure*}[h!]
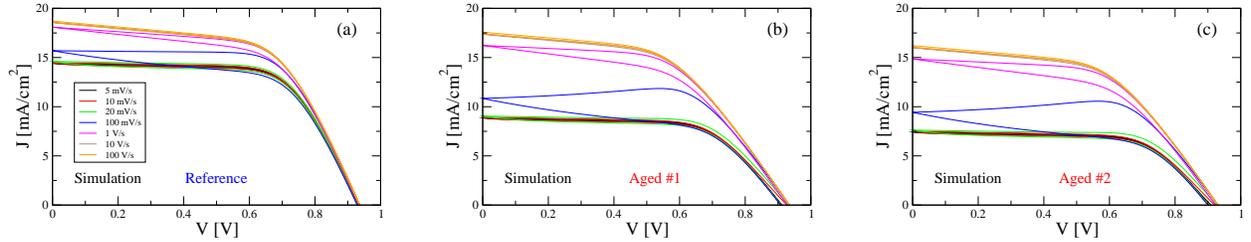

\centering
\includegraphics[width=5cm]{S1a} \hspace*{0.5cm}
\includegraphics[width=5cm]{S1b} \hspace*{0.5cm}
\includegraphics[width=5cm]{S1c}
\caption{Simulated reverse-forward J-V characteristics using DEM, with the parameters indicated in the main text: (a) Reference PSC: $J_{\rm ph} = 14.4$ mA/cm$^2$, $R_{\rm s} = 120$ $\Omega$, $P_0 = 2 P_\infty$; (b) Aged for 1 week: $J_{\rm ph} = 8.9$ mA/cm$^2$, $R_{\rm s} = 200$ $\Omega$, $P_0 = 4 P_\infty$; (c) Aged for 3 weeks: $J_{\rm ph} = 7.4$ mA/cm$^2$, $R_{\rm s} = 200$ $\Omega$, $P_0 = 4 P_\infty$. The aged samples exhibit increased series resistance and a larger initial polarization. Temporary degradation effects were not accounted for.}
\label{sim}
\end{figure*}


\subsection{Temporary degradation induced by bias stress}

\begin{figure*}[h!]
\centering
\includegraphics[width=9cm]{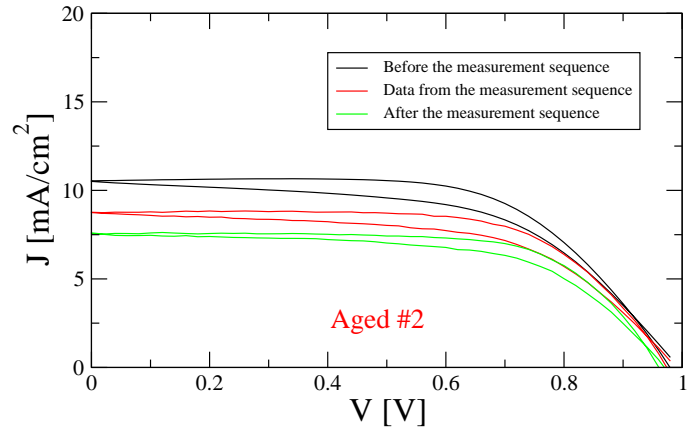} 
\caption{J-V characteristics performed at 20 mV/s showing the temporary degradation induced by performing a sequence of seven J-V scans with the scan rates $\alpha = 10^5, 10^4, 10^3, 10^2, 20, 10, 5 $ mV/s, performed in this order. Typically, the PSCs recover from the bias-stress induced degradation as opposed to irreversible degradation \cite{doi:10.1021/acs.jpclett.6b02375}.}
\label{td}
\end{figure*}

\newpage

\subsection{Inverted hysteresis: low vs. high scan rate}

\begin{figure*}[h!]
\centering
\includegraphics[width=12cm]{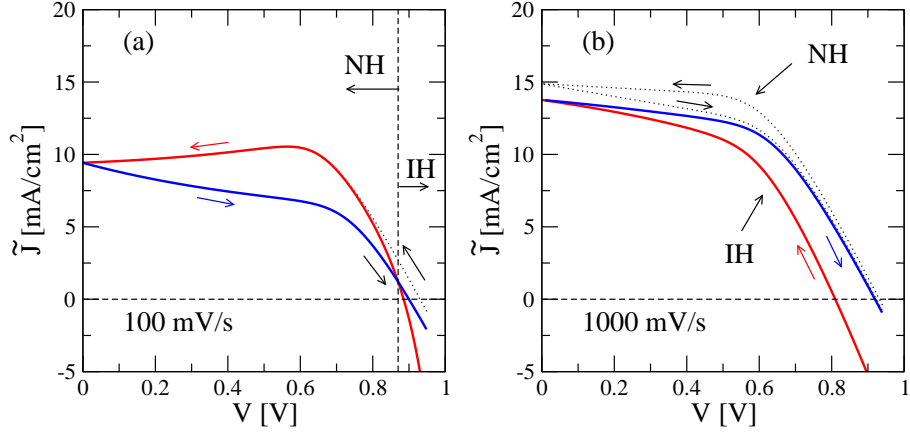} 
\caption{The inverted hysteresis was also observed by pre-conditioning at large positive biases, e.g. using $V_{\rm pol} = 1.4$ V in Figs. 2 and 3 in the main text. We obtain this behavior using DEM \cite{NEMNES2017197,doi:10.1021/acs.jpcc.7b04248}, by including in the simulated J-V characteristics an additional recombination term, i.e. the collected current in the reverse-forward scan becomes $\tilde{J} = J - J_{\rm ph} \exp(-t/\tau_{\rm r})$, where $\tau_{\rm r} = 0.5$ s corresponds to additional recombinations induced by ionic displacement and accumulation at the interfaces, leading to a potential spike in the conduction band \cite{doi:10.1021/acs.jpclett.7b00571}. The current $\tilde{J}$ is represented by solid (red/blue) lines, while $J$, the current in the standard DEM model, is represented by black dotted lines. At a scan rate of 100 mV/s (a), IH appears only near open-circuit, while for a faster scan rate of 1000 mV/s (b), IH becomes dominant for the entire measurement interval. The vertical dashed line in (a) marks the crossing point between the forward and reverse characteristics, separating NH and IH for the scan rate of 100 mV/s. The position of the crossing point depends on both $\tau_r$ and $\alpha$.}
\label{ih}
\end{figure*}


\subsection{PSCs with modified (a) ETL structure and (b) perovskite absorber}

(a) FTO/TiO$_2$-c/TiO$_2$-m/PCBM/CH$_3$NH$_3$PbI$_{2.6}$Cl$_{0.4}$/Spiro/Au

(b) FTO/TiO$_2$-c/TiO$_2$-m/[(CH$_3$NH$_3$)$_{0.95}$(C$_3$N$_2$H$_4$)$_{0.05}$]PbI$_{2.6}$Cl$_{0.4}$/Spiro/Au


\begin{figure*}[h!]
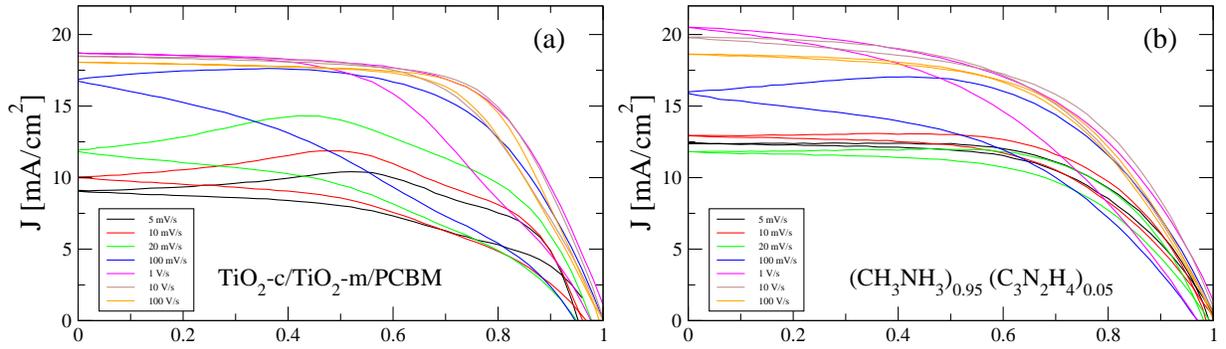

\centering
\includegraphics[width=8cm]{S4a}  
\includegraphics[width=8cm]{S4b} 
\caption{Scan rate dependence of the reverse-forward J-V characteristics of modified PSCs with (a) PCBM additional layer and (b) mixed organic cation perovskite [(CH$_3$NH$_3$)$_{0.95}$(C$_3$N$_2$H$_4$)$_{0.05}$]PbI$_{2.6}$Cl$_{0.4}$. Both types of structures exhibit a similar behavior by varying the bias scan rate: a maximum hysteresis occurs for the intermediate scan rates, the short circuit current increases with the bias scan rate and a current 'bump' is typically visible in reverse characteristics at low to medium scan rates. The hysteresis is minimized for small scan rates (5 mV/s) and for very high ones (100 V/s).}  
\label{pcbm-imi-scan-rate}
\end{figure*}



\begin{figure*}[h!]
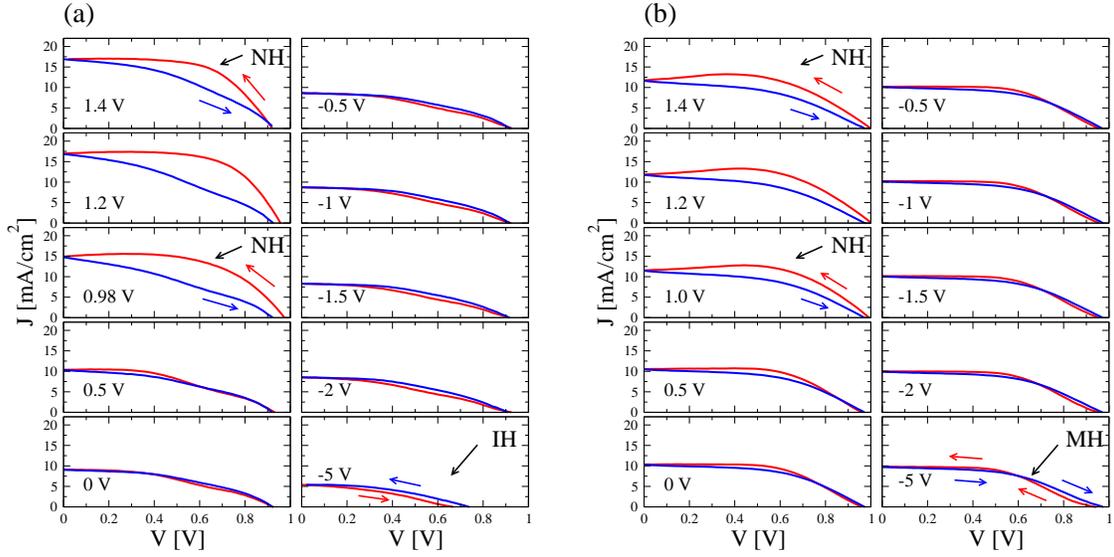

\centering
\includegraphics[width=7cm]{S5a} \hspace*{0.5cm}
\includegraphics[width=7cm]{S5b} 
\caption{Poling effects at a scan rate of 100 mV/s of modified PSCs with (a) PCBM additional layer and (b) mixed organic cation perovskite [(CH$_3$NH$_3$)$_{0.95}$(C$_3$N$_2$H$_4$)$_{0.05}$]PbI$_{2.6}$Cl$_{0.4}$. The bias pre-poling voltage $V_{\rm pol}$ is indicated in each sub-plot, in the range from 1.4 V to -5 V. For large positive poling voltage (e.g. 1 - 1.4 V) the reverse-forward characteristics typically display a large hysteresis, which is reduced as the poling voltage is lowered, yielding almost hysteresis free behavior. For low negative poling voltage (e.g. -5 V) a small inverted hysteresis (a) or a mixed hysteresis, denoted by MH, in (b) is observed.}
\label{pcbm-imi-100mVs}
\end{figure*}



\begin{figure*}[h!]
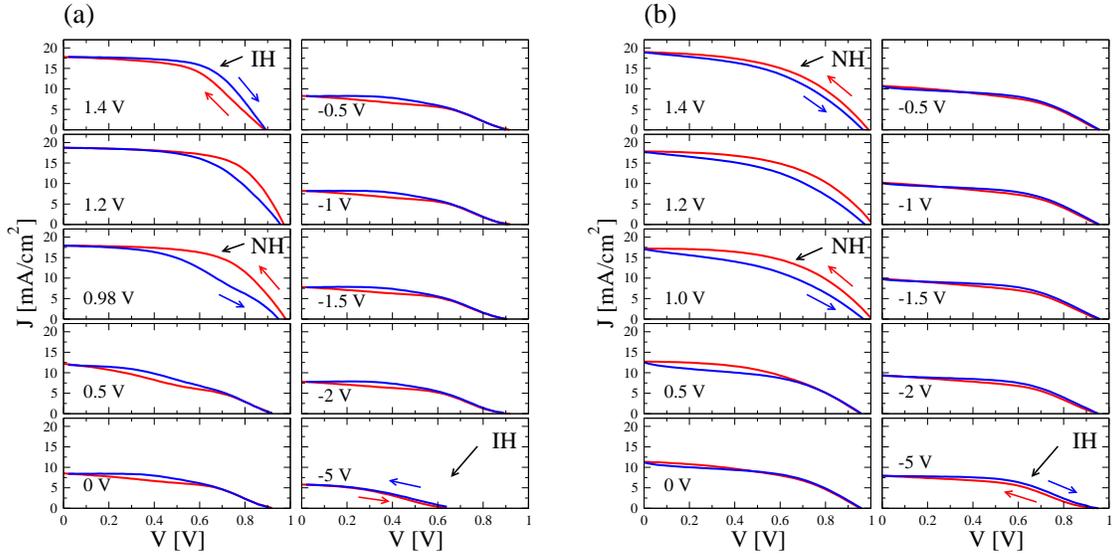

\centering
\includegraphics[width=7.0cm]{S6a} \hspace*{0.5cm}
\includegraphics[width=7.0cm]{S6b} 
\caption{Poling effects at a larger scan rate of 1000 mV/s of modified PSCs with (a) PCBM additional layer and (b) mixed organic cation perovskite [(CH$_3$NH$_3$)$_{0.95}$(C$_3$N$_2$H$_4$)$_{0.05}$]PbI$_{2.6}$Cl$_{0.4}$. As shown in Fig. S3 and Fig. 3 in the manuscript, a large positive poling bias may temporarily enhance the recombinations with a typically shorter time scale $\tau_{\rm r}$. At the high scan rate of 1000 mV/s the effect is present over the entire measurement interval, which, for $V_{\rm pol} = 1.4$ V, is translated into an IH in (a) and a reduced NH, as e.g. compared to $V_{\rm pol}$ = 1.2 V or 1 V in (b). For both types of PSCs, as the poling voltage decreases, the NH is first recovered/enhanced, subsequently turned into a mixed hysteresis and, lastly, into IH.}
\label{pcbm-imi-1000mVs}
\end{figure*}


\end{document}